\documentclass[aps,prl,twocolumn]{revtex4}%
\usepackage{amsfonts}
\usepackage{amsmath}
\usepackage{amssymb}
\usepackage{graphicx}%
\providecommand{\U}[1]{\protect\rule{.1in}{.1in}}

\begin{document}
\preprint{cond-mat/XXX}
\title{Spin resonance in a chiral helimagnet}
\author{Jun-ichiro Kishine}
\affiliation{Department of Basic Sciences, Kyushu Institute of Technology, Kitakyushu
804-8550, Japan}
\author{A. S. Ovchinnikov}
\affiliation{Department of Physics, Ural State University, Ekaterinburg 620083, Russia}

\pacs{PACS number}

\begin{abstract}
It is suggested that marked features of symmetry breaking mechanism and
elementary excitations in chiral helimagnet come up as visible effects in
electron spin resonance (ESR) profile. Under the magnetic field applied
parallel and perpendicular to the helical axis, elementary excitations are
respectively described by the helimagnon associated with rotational symmetry
breaking and the magnetic kink crystal phonon associated with translational
symmetry breaking. We demonstrate how the ESR spectra distinguish these excitations.

\end{abstract}

\maketitle


 In magnetism, chirality means the left- or right-handedness associated with the helical order of magnetic moments. Helimagnetic order can arise from spontaneous symmetry breaking in systems with competing exchange interactions\cite{Yoshimori59} (\lq\lq symmetric\rq\rq helimagnets), or it can be stabilized by the Dzyaloshinkii-Moriya (DM) antisymmetric exchange interaction\cite{Dzyaloshinskii,Moriya60}, which is realized in crystals lacking rotoinversion symmetry (\lq\lq chiral\rq\rq helimagnets). To clarify physical outcome of the chiral spin modulation is of great interest, especially in connection with the symmetry breaking mechanism and the spectrum of elementary excitations which are quite sensitive to the direction of the applied magnetic field.

\begin{figure}[h]
\begin{center}
\includegraphics[width=85mm]{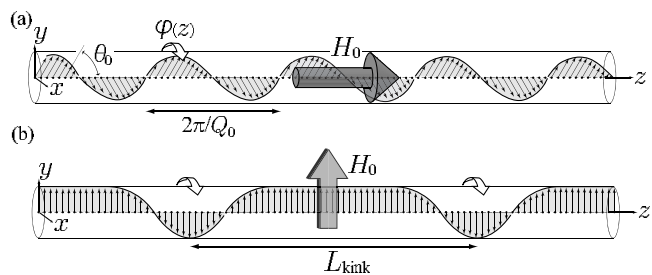}
\end{center}
\caption{(a) Conical and (b) magnetic kink crystal (MKC) states. The helical
axis is $z$-axis.}%
\label{gs}%
\end{figure}

In a chiral helimagnet, under the magnetic field parallel to the helical axis, the ground state (GS) generally changes from planar spiral to conical states [Fig.~\ref{gs}(a)]. The
incommensurate modulation period $2\pi/Q_{0}$ is fixed through $Q_{0}%
=\tan^{-1}(D/J)$, where $D$ and $J$\ are nearest-neighbor DM$\ $interaction
and ferromagnetic exchange interaction
strengths\cite{Dzyaloshinskii64,Izyumov84}. The GS has infinite degeneracy
associated with arbitrary choice of the origin of the phase angle $\varphi
_{0}$. Consequently, the \textit{rotational} symmetry around the helical axis
is spontaneously broken. Then, there appears helimagnetic spin-wave (chiral
helimagnon) mode\cite{Elliot66} as the Nambu-Goldstone (NG) mode, which is
well described in conventional spin wave picture. The chiral helimagnon has
been studied in the context of cubic magnet MnSi\cite{Kataoka87,Maleyev06} and
its peculiar nature has attracted much attention in its own
right\cite{Kirkpatrick}.

On the other hand, under the magnetic field applied perpendicular to the
helical axis, the GS possesses a periodic array of the commensurate (C) and
incommensurate (IC) domains partitioned by discommensurations (DCs), i.e.
the\textit{ internal lattice} which is called magnetic kink crystal (MKC) or
sometimes referred to as chiral soliton
lattice\cite{Dzyaloshinskii64,Izyumov84} is stabilized as shown in
Fig.~\ref{gs}(b). Actually, formation of the MKC state is reported in
CuB$_{2}$O$_{4}$\cite{Rosseli}. This state is also regarded as non-trivial
topological GS. The topological GS in chiral magnet has attracted active
attention from various viewpoints\cite{MnSiSkyrmion}. As the magnetic field
strength increases, the spatial period of MKC lattice, $L_{\text{kink}}$,
increases and finally goes to infinity at the critical field strength.
Recently, we showed that this internal lattice exhibits mutual sliding which
may be experimentally detectable\cite{BKO08,BKO09}. In this case, the GS has
infinite degeneracy associated with arbitrary choice of the center of mass
position. Consequently, the \textit{translational} symmetry along the helical
axis is spontaneously broken. Then, the elementary excitations are described
by \textquotedblleft phonon\textquotedblright mode of correlated kinks. What
is interesting is that we can control the size of the first Brillouin zone of
the MKC lattice upon changing the magnetic field strength.

The elementary excitations in the kink crystal state was first investigated by
Sutherland\cite{Sutherland73}. He considered the\ sine-Gordon model for a
single scalar field corresponding to the tangential $\varphi$-mode of the
planar XY spins and found that the elementary excitations consist of the
acoustic and optical bands separated by the energy gap. The acoustic band is
formed out of correlated translations of the individual kinks and corresponds
to gapless NG bosons. The optical band corresponds to renormalized
Klein-Gordon bosons. In chiral helimagnet, we need to take account of not only
the $\varphi$-mode but the longitudinal $\theta$-mode ($\theta$ is an angle
between the spin vector and the helical axis). In previous works\cite{BKO08},
we pointed out that the $\theta$-mode acquires an energy gap originating from
the DM interaction.

Then, natural question arises as to whether the helimagnon and MKC phonon have
observable consequences for the magnetic response using ESR technique. In this
paper, we demonstrate how the symmetry breaking patterns and the elementary
excitations come up in the ESR signals.

In the ESR experiment, the static magnetic field $\mathbf{H}_{0}$\ is applied
to cause Larmor precession of magnetic spins. Then supplying electromagnetic
energy carried by microwave radiation, resonant absorption occurs at the
precession frequency. The microwave is described as the \textit{uniform}
oscillating magnetic field (rf field) $\mathbf{h}(t)$ polarized in the
direction perpendicular to $\mathbf{H}_{0}$ (Faraday configuration). The rf
field gives rise to the Zeeman coupling with spin, $\mathcal{H}_{Z}%
=-\mathbf{H}(t)\cdot\mathbf{S}_{0}$, where $\mathbf{H}(t)=g_{e}\mu
_{B}\mathbf{h}(t)$ ($g_{e}$ is the electron's g-factor and $\mu
_{B}$ is the Bohr magneton) and $\mathbf{S}_{0}$ is the uniform ($q=0$) component of the spin variable. For $\mathbf{H}(t)=H_{\mu}\mathbf{\hat{e}}^{\mu}\cos\left(\omega t\right)  $, the ESR spectrum (absorbed energy per unit
time) is given by, $\mathcal{Q}\left(  \omega\right)  =\omega H_{\mu}^{2}%
\chi_{\mu\mu}^{\prime\prime}\left(  \omega\right)  /2,$ where $\mathbf{\hat{e}}^{\mu}$ with $\mu=x,y,z$ denotes the unit vector along $x$, $y$, and $z$ axis [Fig.~1], respectively, and $\omega$ is a microwave frequency. The imaginary part 
of the dynamical susceptibility, $\chi_{\mu\nu}^{\prime\prime}\left(
\omega\right)  =\left(  1-e^{-\hbar\omega/k_{\rm{B}}T}\right)  C_{\mu\nu}\left(
\omega\right)  /2$, is related to the correlation function $C_{\mu\nu}\left(
\omega\right)  =\left\langle S_{0}^{\mu}\left(  \omega\right)  S_{0}^{\nu
}\right\rangle $ through the fluctuation-dissipation theorem. In quantum
mechanical language, the Lamor precession corresponds to equally spaced Zeeman
splitting of the energy levels. Because of the equal spacing of the quantum
energy levels, the quantum-classical correspondence exactly holds and the
classical frequency is equal to quantum one as far as we consider Gaussian fluctuations.

First, we consider the case where the magnetic field is applied parallel to
the helical axis ($z$-axis) and the rf field is polarized along the $y$-axis.
Then, the elementary excitations are described as spin waves over the conical
magnetic structure. A quantized spin wave is helimagnon. Then, the ESR spectrum
is given by $\mathcal{Q}_{\text{hmag}}\left(  \omega\right)  =\omega H_{y}%
^{2}\chi_{yy}^{\prime\prime}\left(  \omega\right)  /2.$ To compute $\chi
_{yy}^{\prime\prime}\left(  \omega\right)  $, we{ assume that the magnetic
atoms form }{a three dimensional}{ lattice and a uniform ferromagnetic
coupling exists between the adjacent chains to stabilize the long-range order.
Then, the Hamiltonian is interpreted as an effective one-dimensional model
based on the interchain mean field picture and is written as},
\begin{gather}
\mathcal{H}=-{\frac{\tilde{J}}{2}}%
{\displaystyle\sum\limits_{j}}
[e^{iQ_{0}c}S_{j}^{+}S_{j+1}^{-}+e^{-iQ_{0}c}S_{j}^{-}S_{j+1}^{+}]\nonumber\\
-J%
{\displaystyle\sum\limits_{j}}
S_{j}^{z}S_{j+1}^{z}+K_{\bot}%
{\displaystyle\sum\limits_{j}}
(S_{j}^{z})^{2}-\mathbf{H}_{0}\cdot%
{\displaystyle\sum\limits_{j}}
\mathbf{S}_{j}, \label{Hamiltonian}%
\end{gather}
where $\mathbf{S}_{j}$ represents a spin located at the $j$-th site along the
helical axis ($z$-axis) and $S_{j}^{\pm}=S_{j}^{x}\pm iS_{j}^{y}.$ The
mono-axial DM vector is $\mathbf{D}=D\mathbf{\hat{e}}^{z}$ and $\tilde
{J}=|J+iD|=\sqrt{J^{2}+D^{2}}$. The lattice constant is $c$. We include the
easy-plane anisotropy with strength $K_{\bot}$. For $H_{0}=0$, the planar
helical structure is stable under the condition $K_{\bot}/J>1-\sqrt
{1+(D/J)^{2}}$ which is assumed to be satisfied. For $0<H_{0}<H_{0c}%
=2S(\tilde{J}-J+K_{\bot})$, the GS is described by $S_{j}^{\pm}=Se^{\pm
i\left(  Q_{0}z_{j}+\varphi_{0}\right)  }\sin\theta$, where the cone angle is
given by $\theta=\theta_{0}=\cos^{-1}[H_{0}/\{2S(\tilde{J}-J+K_{\bot})\}].$

\begin{figure}[ptb]
\begin{center}
\includegraphics[width=85mm]{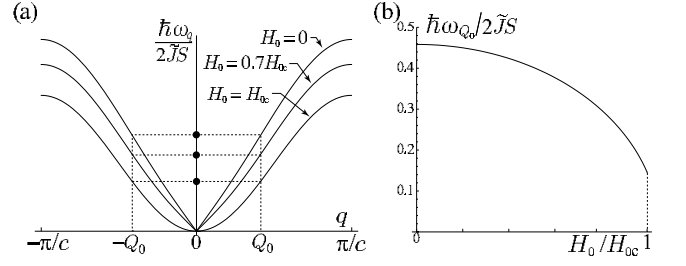}
\end{center}
\caption{(a) Helimagnon dispersions for $H_{0}/H_{0c}=0,$ $0.7,$ and $1.$. We
took $D/J=0.5$ and $K_{\bot}/J=2$.\ Black dots indicate the location of the
resonance energies. (b) Field dependence of the resonance energy as a function
of $H_{0}/H_{0c}$. }%
\label{spin_wave}%
\end{figure}

To obtain the spin wave spectrum, we rotate the basis frame of the crystal
coordinate $\{\mathbf{\hat{e}}^{+},\mathbf{\hat{e}}^{-},\mathbf{\hat{e}}%
^{z}\}$ to the basis frame of the local coordinate $\{\mathbf{\hat{e}}_{j}%
^{+},\mathbf{\hat{e}}_{j}^{-},\mathbf{\hat{e}}_{j}^{z}\}$ where the direction
of $\mathbf{\hat{e}}_{j}^{z}$ points to the equilibrium spin direction at the
$j$-th site\cite{Nagamiya-review}. In the spirit of conventional spin-wave
approximation, we obtain the spectrum,%
\begin{equation}
\frac{\hbar\omega_{q}}{2\tilde{J}S}=\sqrt{\left[  1-\cos(qc)\right]
[\bar{\lambda}-\bar{\gamma}\cos(qc)]},
\end{equation}
where $q$ is a wave number, $\bar{\lambda}=1+(K_{\bot}/\tilde{J})\sin^{2}\theta_{0}$ and
$\bar{\gamma}=(J/\tilde{J})\sin^{2}\theta_{0}+\cos^{2}\theta_{0}$. This result
reduces to the one obtained by Kataoka\cite{Kataoka87} and
Maleyev\cite{Maleyev06} using the continuum approximation ($q\rightarrow0$
limit). In Fig.~\ref{spin_wave}(a), we show the helimagnon dispersion for
$H_{0}=0,$ $0.7H_{0c},$ and $H_{0c}$. Upon increasing the field, linear
dispersions for $0\leq H_{0}<H_{0c}$ continuously crosses over to the
quadratic dispersion $\hbar\omega_{q}=2\tilde{J}S\left(  1-\cos q\right)
$\ at $H_{0}=H_{0c}$. The Goldstone mode at $q=0$ corresponds to the rigid
rotation of the whole helix. For $H_{0}\geq H_{0c}$, the equilibrium state is
forced-ferromagnetic state and the spin wave spectrum acquire the
field-induced gap. 

Now, it is straightforward to obtain the helimagnon resonance spectrum,
\begin{gather}
\mathcal{Q}_{\text{hmag}}\left(  \omega\right)  =\frac{\pi S}{8}\omega
H_{y}^{2}\delta\left(  \omega-\omega_{Q_{0}}\right) \nonumber\\
\times\lbrack(u_{Q_{0}}^{+}+u_{Q_{0}}^{-})^{2}+\cos^{2}\theta_{0}(u_{Q_{0}%
}^{+}-u_{Q_{0}}^{-})^{2}], \label{helimagresonance}%
\end{gather}
where $u_{Q_{0}}^{\pm}=\sqrt{\left(  P/\omega_{Q_{0}}\pm1\right)  /2}$ and
$P=S|2\tilde{J}+K_{\bot}\sin^{2}\theta_{0}-J\{1+\cos^{2}\theta_{0}%
+(J/\tilde{J})\sin^{2}\theta_{0}\}|$. Note that the external uniform field
couples to the $q=\pm Q_{0}$ component of the spin wave excitation, since the
field is seen in the local frame as spatially rotating field with modulation
wave-number $Q_{0}$. Consequently, we have a single branch of resonance
energy, as shown in Fig.~\ref{spin_wave}(b). As we shall see, this situation
drastically changes in the case of the MKC phonon resonance.

ESR signal in chiral helimagnet MnSi was reported by Date \textit{et
al.}\cite{Date77}. At that time, however, they adopted the formula obtained by
Yoshimori\cite{Yoshimori59} and Cooper \textit{et al.}\cite{Cooper62} for
symmetric helimagnetic structure stabilized by frustration among the exchange
interactions\cite{Yoshimori59}. In the case of symmetric helimagnet, the spin
wave dispersion exhibits dips at $q=\pm Q_{0}$ and the corresponding energy
gaps vanish for $K_{\bot}=0$\cite{Nagamiya-review}. There are no such
additional dips in chiral helimagnon spectrum. We see, however, it may not be
easy to distinguish the spin wave spectra of chiral helimagnet from those of
symmetric helimagnet simply by ESR profile, because both cases give apparently
quite similar field dependence of the resonance energies as shown in
Fig.~\ref{spin_wave}(b).

Next, we consider the MKC phonon resonance when the magnetic field is applied
perpendicular to the helical axis ($y$-axis) and the rf field is polarized
along the $z$-axis. The MKC state is described in terms of the slowly varying
polar angles $\theta(z)$ and $\varphi(z)$. The vector spin density is defined
by $\mathbf{S}\left(  z\right)  =\sum_{j}\mathbf{S}_{j}\delta\left(
z-z_{j}\right)  =\left(  \sin\theta(z)\cos\varphi(z),\sin\theta(z)\sin
\varphi(z),\cos\theta(z)\right)  .$ Then, minimizing the continuum version of
the Hamiltonian (\ref{Hamiltonian}), we obtain the MKC state as a stationary
state described by $\theta=\pi/2$ and $\cos[\varphi_{0}(z)/2]=\mathrm{sn}%
(2Kz/L_{\text{kink}}),$ where $L_{\text{kink}}={8K}E/\pi Q_{0}$ is the period
of the MKC lattice. $K$ and $E$ denote the elliptic integrals of the first and
second kind, respectively, with the elliptic modulus $\kappa$ ($0\leq
\kappa\leq1$). \lq\lq$\,$\textrm{sn}\rq\rq is Jacobi-\textrm{sn }function. The
IC to C transition occurs at the critical field strength $H_{0}^{\ast}/{{{J}%
S}}=\left(  {\pi Q_{0}/4}\right)  ^{2}$ at which $L_{\text{kink}}$ diverges.
The elliptic modulus $\kappa$ is determined by the condition $\sqrt
{H_{0}/H_{0}^{\ast}}=\kappa/E\left(  \kappa\right)  .$ The IC to C transition
in chiral magnet is actually reported in real materials\cite{KIY}. For
example, in the case of Cr$_{1/3}$NbS$_{2}$\cite{Miyadai}, $H_{0}^{\ast}$
takes values from about 1 to 1.4 kOe and in the case of CuB$_{2}$O$_{4}%
$\cite{Kousaka}, from 0.5 to 10kOe depending on temperatures.

In this case, the rf field couples with $S^{z}(z,t)=S\cos
\theta(z,t)$ and ESR spectrum is given by $\mathcal{Q}_{\text{ph}%
}\left(  \omega\right)  =\omega H_{z}^{2}\chi_{zz}^{\prime\prime}\left(
\omega\right)  /2.$ To compute $\chi_{zz}^{\prime\prime}\left(  \omega\right)
$, we need the explicit form of the propagating mode $S^{z}(z,t)\simeq
-Su(z,t)$ where $u({z},t)=\theta(z,t)-\pi/2$\ describes small fluctuation
around the MKC state. Although full description should include the $\varphi
$-mode, the rf field couples to only $\theta$-mode and it is enough to
consider the $\theta$-mode only. By using the mode expansion for $u(z,t)$, we
set up the vibrational Hamiltonian given as collections of harmonic
oscillators\cite{BKO08}. The explicit form of the quantized phonon wave
function is given by
\begin{equation}
u(z,t)=\sum_{q}\sum_{n=-\infty}^{\infty}\left[  \dfrac{U_{n}}{\sqrt
{2\omega_{q}}}e^{-i(q-nG_{\text{MKC}})z+i\omega_{q}t}b_{q}^{\dagger
}+\text{h.c.}\right]  , \label{phonon}%
\end{equation}
where $b_{q}^{+}\left(  b_{q}\right)  $ are the phonon creation (annihilation)
operators. The crystal-momentum $q$ and the eigenfrequency $\omega_{q}$ are
expressed in terms of a real parameter $a$\ running over $-K^{\prime}<a\leq
K^{\prime}$, where $K^{\prime}$ is the complete elliptic integral of the first
kind with the complementary modulus $\kappa^{\prime}$. For the acoustic
branch, $q=l_{0}^{-1}[Z(a,\kappa^{\prime})+\pi a/(2KK^{\prime})]$ and
$\hbar\omega_{q}=\varepsilon_{0}\sqrt{\Delta^{2}+\kappa^{\prime2}%
\,\mathrm{sn\,}^{2}\left(  a,\kappa^{\prime}\right)  /2}$. For the optical
branch, $q=l_{0}^{-1}[Z(a,\kappa^{\prime})+\pi a/(2KK^{\prime})+\mathrm{{dn\,}%
}(a,\kappa^{\prime})\mathrm{{cs\,}}(a,\kappa^{\prime})]$ and $\hbar\omega
_{q}=\varepsilon_{0}\sqrt{\Delta^{2}+\mathrm{sn\,}^{-2}\left(  a,\kappa
^{\prime}\right)  /2}$. $Z(a,\kappa^{\prime})$ is the elliptic zeta-function.
We here introduced the characteristic length and energy units $l_{0}%
=L_{\text{kink}}/2K={4}E/\pi Q_{0}\simeq Q_{0}^{-1}$ and $\varepsilon
_{0}=JS^{2}c/l_{0}=JS^{2}\pi Q_{0}c/4E\simeq DS^{2}c$, respectively. It is
essential that the energy gap $\Delta=\sqrt{8E/\pi-2}$ opens at $q=0$, because
of the existence of the DM interaction. The Fourier coefficients $U_{n}$ can
be computed by performing contour integral of the real space wave function
given in Ref. \cite{BKO08}. To obtain the ESR spectrum, we need $U_{0}%
=\pi/(2K\sqrt{\kappa^{\prime}})$ and $U_{n}=-i^{-1}\vartheta_{1}^{\prime
-1}\vartheta_{1}(i\pi a_{n}/2K)/\sinh[\pi a_{n}/(2K)-n\pi K^{\prime}/K]$
for\ $n\neq0$ [$a_{n}$ is determined by the resonance condition
(\ref{resonance}) given below]. $\vartheta_{1}$ is the Jacobi theta function
and $\vartheta_{1}^{\prime}=\vartheta_{1}^{\prime}(0)$. Since $\vartheta
_{1}(i\pi a/2K)$ is purely imaginary, all $U_{n}$ are real.

The reciprocal lattice constant of the MKC lattice is given by%
\begin{equation}
G_{\text{MKC}}=\frac{2\pi}{L_{\text{kink}}}=\frac{\pi^{2}}{4KE}Q_{0}.
\label{rlc}%
\end{equation}
The first Brillouin zone of the MKC lattice is $\left\vert q\right\vert \leq
G_{\text{MKC}}/2$ and the energy gap between the acoustic and optical branches
opens at the zone boundary. As limiting forms, we have $\hbar\omega_{q}%
\simeq\varepsilon_{0}\sqrt{\Delta^{2}+v^{2}q^{2}}$ ($v$ is constant)\ for
$q\ll G_{\text{MKC}}/2$ and $\hbar\omega_{q}\simeq\varepsilon_{0}\left\vert
ql_{0}\right\vert /\sqrt{2}$ for $G_{\text{MKC}}/2\ll q.$ Now, we are ready to
understand the ESR by the MKC phonon. Since the rf field along the $z$-axis
carries the wave number $q=0$, the resonant absorption is caused by the MKC
phonon modes with a series of special wave numbers%
\begin{equation}
q=q_{n}=nG_{\text{MKC}}. \label{resonance}%
\end{equation}
The correlation function can be easily computed by using Eq.(\ref{phonon}) and
we obtain the ESR absorption spectrum,%
\begin{equation}
\mathcal{Q}_{\text{ph}}\left(  \omega\right)  =\frac{\pi\omega}{4}H_{z}%
^{2}S^{2}%
{\displaystyle\sum\limits_{n=0}^{\infty}}
\frac{{\left\vert U_{n}\right\vert }^{2}}{\omega_{n}}\delta\left(
\omega-\omega_{n}\right)  , \label{phonon_resonance}%
\end{equation}
where $\omega_{n}=\omega_{q_{n}}$. This expression together with $U_{n}$ and
$\omega_{n}$ complete a closed formula for the MKC phonon resonance. For
$n=0$, the bottom of the acoustic branch ($q=0$ and $a=0$) gives $\omega
_{0}=\Delta$. For $n\geq1$, the optical branch contributes to the resonance.

\begin{figure}[t]
\begin{center}
\includegraphics[width=83mm]{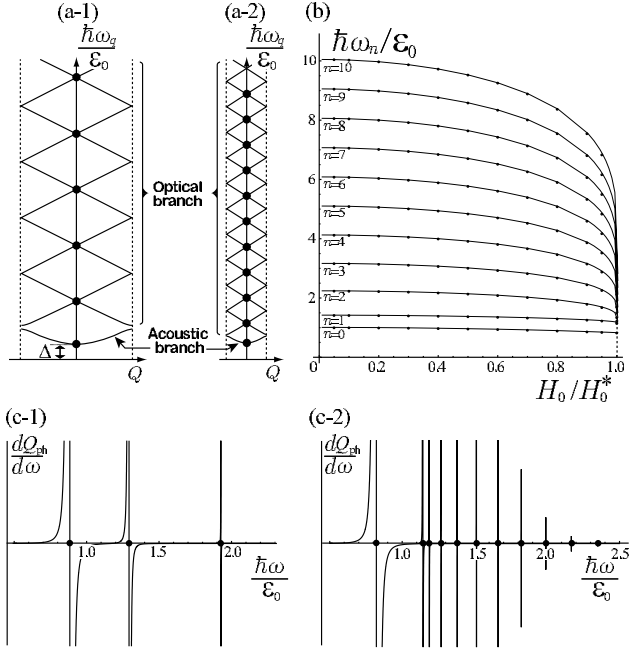}
\end{center}
\caption{Energy dispersion of the MKC phonon in the reduced zone scheme for
(a-1) smaller and (a-2) larger magnetic field strengths. The vertical broken
lines indicate the Brillouin zone boundaries $q=\pm G_{\text{MKC}}/2.$ (b)
Resonance energy $\omega_{{n}}$ for $n=0$ to $n=10$ as functions of
$H_{0}/H_{0}^{\ast}$. \ We took $D/J=0.5$ and $K_{\bot}/J=0$.\ \ The
derivative absorption $d\mathcal{Q}_{\text{ph}}/d\omega$ for (c-1)
$H_{0}/H_{0}^{\ast}=0.8$ and (c-2) $H_{0}/H_{0}^{\ast}=1-10^{-8}$. In (a-1),
(a-2), (c-1), and (c-2), black dots indicate the location of the resonance
energies.}%
\label{MKCresonancefig}%
\end{figure}

As the magnetic field increases from zero to $H_{0}^{\ast}$, $G_{\text{MKC}}%
$\ decreases from $Q_{0}$ to zero. On the other hand, the original atomic
lattice constant $c$ gives natural cutoff and the atomic Brillouin zone
boundary $\pm2\pi/c$ irrespective of the external magnetic field. Usually,
$2\pi/Q_{0}\simeq10c-100c$ and therefore $G_{\text{MKC}}$ is much smaller than
the atomic zone boundary $2\pi/c.$ In Fig.~\ref{MKCresonancefig}(a), we
schematically depict that the distribution of the resonance energy levels
becomes more and more dense upon increasing the magnetic field strength. In
Fig.~\ref{MKCresonancefig}(b), we show the resonance energies $\omega_{n}$ for
$n=0$ to $n=10$ as functions of $H_{0}/H_{0}^{\ast}$. To obtain the result
presented here, we first numerically solve Eq. (\ref{resonance}) in terms of
the parameter $a$, and then compute the corresponding frequency $\omega_{n}$,
which completes evaluation of Eq. (\ref{phonon_resonance}). In
Fig.~\ref{MKCresonancefig}(c-1), we show the derivative absorption
$d\mathcal{Q}_{\text{ph}}\left(  \omega\right)  /d\omega$ for $H_{0}%
/H_{0}^{\ast}=0.8$. The delta function is replaced by $\delta(\omega
)=\pi^{-1}\epsilon/(\omega^{2}+\epsilon^{2})$ with $\epsilon=10^{-4}$.
Although the weight ${\left\vert U_{n}\right\vert }^{2}$\ rapidly decays for
higher order resonances, the peak structure becomes visible by taking the derivative.

Of special interest is the region in the vicinity of the IC-C transition,
where the distribution of the resonance levels are quite dense. In
Fig.~\ref{MKCresonancefig}(c-2), we show the case for $H_{0}/H_{0}^{\ast
}\rightarrow1$. We see that a series of many densely spaced resonance lines
appear. Using the relation $K(\kappa)\simeq\log(4/\sqrt{1-\kappa^{2}})$ and
$E(\kappa)\simeq1$ which hold for $\kappa\lesssim1$, we have $\kappa
\simeq\sqrt{H_{0}/H_{0}^{\ast}}$ and therefore obtain the asymptotic form of
the resonance frequencies\ for large $n$,%
\begin{equation}
\frac{\hbar\omega_{n}}{\varepsilon_{0}}\simeq n\frac{\pi}{K}\simeq\frac{n\pi
}{\log\left(  4/\sqrt{1-H_{0}/H_{0}^{\ast}}\right)  }. \label{af}%
\end{equation}
Conversely, a series of resonance fields for large $n$ are given by
$H_{0n}/H_{0}^{\ast}\simeq1-16\exp\left(  -2\pi n\varepsilon_{0}/\hbar
\omega\right)  $ for a fixed frequency $\omega$. Our energy unit
$\varepsilon_{0}\simeq DS^{2}c$ usually amounts to $J/100-J/10$, corresponding
to 1meV to 10meV in energy scales. These energy scales correspond to microwave
frequencies in THz region (quantitative detail depends on $\varepsilon_{0}$).
So, our effects should be detectable in submillimeter wave ESR
measurements\cite{Ajiro}.

We stress that the MKC phonon resonance never occurs in the symmetric
helimagnet due to energetic frustration\cite{Yoshimori59}, where not the MKC
but the \textquotedblleft fan\textquotedblright\ structure is stabilized under
the field perpendicular to the helical axis\cite{Nagamiya-review,Cooper62}.
Physical background behind this difference is that in chiral helimagnet the
crystallographic chirality plays a role of \textquotedblleft topological
protectorate\textquotedblright\ for the MKC lattice state to appear as the
stable GS.

In basic physical ideas, the effects we proposed here is one of examples to
detect spin dynamics of phase modulated states by polarized probes such as
inelastic neutron beam or X-ray. For example, neutron beams can probe the MKC
phonon mode via the differential cross section $d^{2}\sigma/d\Omega
d\omega\varpropto(1-\mathbf{\hat{k}}_{z}^{2})\left\langle S_{\mathbf{k}}%
^{z}(\omega)S_{-\mathbf{k}}^{z}\right\rangle $ where $\omega$ and $\mathbf{k}$
are respectively frequency and scattering wave-number of the
neutron\cite{Izyumov-Laptev86}. Then, the scattering event occurs when both
momentum conservation, $k_{z}=q-nG_{\rm{MKC}}$, and the energy conservation, $\omega
=\pm\omega_{q}$, are satisfied, where $q$ is the MKC phonon wave-number. The
polarized X-ray beam may also detect the MKC state via the generalized
spin-orbit coupling between the spin magnetic moment and X-ray. These topics
will be treated separately in a forthcoming paper.

Finally, we make theoretical comments on the MKC state. The MKC apparently
seems to be {\it one}-dimensional object which is fragile against three dimensional couplings.
However, it is not necessary to worry about this. Many features in physics of incommensurate
magnets may be understood based on the Ginzburg-Landau free energy with a non-uniform order
parameter as a function of {\it three}-dimensional coordinates. Then one should select a solution minimizing the functional that corresponds to 
the modulated phase. As a result of this analysis, we find a structure with a modulation along
{\it one} axis in the crystal is easily stabilized. In such a case, it is enough to take into account
the invariant involving derivatives with respect to one coordinate ($z$-coordinate in the present case). 
More rigorously speaking,  we need to exclude a possibility that we have 
a structure with multiple modulation vectors in a single crystallographic domain.
But it is known that the realization of this kind of structure 
is hard to occur (see for example, Ref.\cite{Izyumov_Syromyatnikov}). 
This is the reason why we can safely start with 
the effective one-dimensional model as we did in this paper.


\begin{acknowledgments}
J.~K. acknowledges Grant-in-Aid for Scientific Research (A)(No.~18205023) and
(C) (No.~19540371) from the Ministry of Education, Culture, Sports, Science
and Technology, Japan.
\end{acknowledgments}

\end{document}